\pdfoutput=1 
\documentclass[a4paper]{article}
\usepackage[a4paper]{geometry}

\usepackage{tikz}
\usepackage[T1]{fontenc}
\usepackage{agda}

\begin{code}[hide]%
\>[0]\AgdaKeyword{module}\AgdaSpace{}%
\AgdaModule{FlipperLAFI}\AgdaSpace{}%
\AgdaKeyword{where}\<%
\\
\\[\AgdaEmptyExtraSkip]%
\>[0]\AgdaKeyword{open}\AgdaSpace{}%
\AgdaKeyword{import}\AgdaSpace{}%
\AgdaModule{Prelude}\AgdaSpace{}%
\AgdaKeyword{hiding}\AgdaSpace{}%
\AgdaSymbol{(}\AgdaFunction{flip}\AgdaSymbol{;}\AgdaSpace{}%
\AgdaFunction{uncurry}\AgdaSymbol{)}\<%
\end{code}

\usepackage[font=small,labelfont=bf]{caption}
\usepackage{j}
\usepackage{inconsolata}
\usepackage{stmaryrd}
\usepackage{graphicx}
\usepackage{polytable}

\DeclareRobustCommand\bdiv{%
  \nonscript\mskip-\medmuskip\mkern5mu%
  \mathbin{\operator@font div}\penalty900\mkern5mu%
  \nonscript\mskip-\medmuskip}

\title{Verified Reversible Programming for\\
Verified Lossless Compression}

\author{James Townsend and Jan-Willem van de Meent\\
University of Amsterdam}

\date{}

\begin{document}

\maketitle

\begin{abstract}
Lossless compression implementations typically contain two programs,
an encoder and a decoder, which are required to be inverse to one
another. We observe that a significant class of compression methods,
based on asymmetric numeral systems (ANS), have shared structure
between the encoder and decoder---the decoder program is the `reverse'
of the encoder program---allowing both to be simultaneously specified
by a single, reversible function. To exploit this, we have implemented
a small reversible language, embedded in Agda, which we call
`Flipper' (available at \jurl{github.com/j-towns/flipper}). Agda supports
formal verification of program properties, and the compiler for our reversible
language (which is implemented as an Agda macro), produces not just an
encoder/decoder pair of functions but also a proof that they are inverse to one
another. Thus users of the language get formal verification `for free'. We give
a small example use-case of Flipper in this paper, and plan to publish a full
compression implementation soon.
\end{abstract}
\vspace{5mm}
It has been known since the work of Claude Shannon in the late 1940s
that there is a deep connection between probabilistic modelling
and lossless data compression. In \cite{shannon1948}, Shannon showed
that the best achievable compressed size for random data is equal to
the expected negative log-probability, a quantity which he called the
`entropy'.  Practical approaches to lossless compression work by
assuming a model \(P\) over input data \(x\), and compressing \(x\) to
a size close to the negative log-probability \(-\log_2 P(x)\), using a
coding method such as Huffman coding \cite{huffman1952}, arithmetic
coding \cite{moffat1998}, or asymmetric numeral systems (ANS)
\cite{duda2009}.
Recently, a number of papers have shown that modern deep generative
models can be used to achieve state-of-the-art compression rates
\cite{townsend2019,%
kingma2019a,townsend2020,hoogeboom2019,berg2021,ruan2021,zhang2021,%
kingma2021}. These recent works all use ANS, a method which, due to
its simplicity and superior performance, has also become ubiquitous in
production compression systems.\footnote{A list is maintained at
\jurl{encode.su/threads/%
2078-List-of-Asymmetric-Numeral-Systems-implementations}, which
includes various ANS programs with state-of-the-art performance.}

Our work on this paper began with an observation: in ANS-based
methods, the encoder and decoder programs appear to have the same
structure.  Indeed, we realised that it made sense to view them as
different interpretations of \emph{the same} program, one `doing' the
program and the other `undoing' it. Following this idea to its logical
conclusion, we have implemented a small, \emph{reversible} domain
specific language, called Flipper, embedded in Agda \cite{norell2007},
via an Agda macro. The potential advantages of using Flipper are
significant---the amount of user code is reduced, and accidental
violation of the inverse property (a common cause of bugs in lossless
compression) is no longer possible.

Flipper programs inhabit a record type of bijections between two sets
\AgdaBound{A} and \AgdaBound{B}:\footnote{%
We would have preferred the names `do' and `undo' to `apply' and
`unapply', but unfortunately `do' is a reserved keyword in Agda.
}\nopagebreak%
\begin{code}%
\>[0]\AgdaKeyword{record}\AgdaSpace{}%
\AgdaOperator{\AgdaRecord{\AgdaUnderscore{}\ensuremath{\leftrightarrow}\AgdaUnderscore{}}}\AgdaSpace{}%
\AgdaSymbol{(}\AgdaBound{A}\AgdaSpace{}%
\AgdaBound{B}\AgdaSpace{}%
\AgdaSymbol{:}\AgdaSpace{}%
\AgdaPrimitive{Set}\AgdaSymbol{)}\AgdaSpace{}%
\AgdaSymbol{:}\AgdaSpace{}%
\AgdaPrimitive{Set}\AgdaSpace{}%
\AgdaKeyword{where}\<%
\\
\>[0][@{}l@{\AgdaIndent{0}}]%
\>[2]\AgdaKeyword{field}\<%
\\
\>[2][@{}l@{\AgdaIndent{0}}]%
\>[4]\AgdaField{apply}\AgdaSpace{}%
\AgdaSymbol{:}\AgdaSpace{}%
\AgdaBound{A}\AgdaSpace{}%
\AgdaSymbol{\ensuremath{\rightarrow}}\AgdaSpace{}%
\AgdaBound{B}\<%
\\
\>[4]\AgdaField{unapply}\AgdaSpace{}%
\AgdaSymbol{:}\AgdaSpace{}%
\AgdaBound{B}\AgdaSpace{}%
\AgdaSymbol{\ensuremath{\rightarrow}}\AgdaSpace{}%
\AgdaBound{A}\<%
\\
\>[4]\AgdaField{prfa}\AgdaSpace{}%
\AgdaSymbol{:}\AgdaSpace{}%
\AgdaSymbol{\ensuremath{\forall}}\AgdaSpace{}%
\AgdaBound{a}\AgdaSpace{}%
\AgdaSymbol{\ensuremath{\rightarrow}}\AgdaSpace{}%
\AgdaField{unapply}\AgdaSpace{}%
\AgdaSymbol{(}\AgdaField{apply}\AgdaSpace{}%
\AgdaBound{a}\AgdaSymbol{)}%
\>[43]\AgdaOperator{\AgdaDatatype{\ensuremath{\equiv}}}\AgdaSpace{}%
\AgdaBound{a}\<%
\\
\>[4]\AgdaField{prfb}\AgdaSpace{}%
\AgdaSymbol{:}\AgdaSpace{}%
\AgdaSymbol{\ensuremath{\forall}}\AgdaSpace{}%
\AgdaBound{b}\AgdaSpace{}%
\AgdaSymbol{\ensuremath{\rightarrow}}\AgdaSpace{}%
\AgdaField{apply}%
\>[31]\AgdaSymbol{(}\AgdaField{unapply}\AgdaSpace{}%
\AgdaBound{b}\AgdaSymbol{)}\AgdaSpace{}%
\AgdaOperator{\AgdaDatatype{\ensuremath{\equiv}}}\AgdaSpace{}%
\AgdaBound{b}\<%
\end{code}
The Flipper macro takes a user implementation of
\AgdaField{apply} in
\AgdaBound{A}
\AgdaSymbol{\ensuremath{\rightarrow}}
\AgdaBound{B},
and compiles it into an inhabitant of
\AgdaBound{A}
\AgdaFunction{\ensuremath{\leftrightarrow}}
\AgdaBound{B}, complete with an implementation of \AgdaField{unapply}
and the two necessary correctness proofs.

The compiler only accepts functions expressed in an `obviously'
reversible form, where the reverse interpretation can be seen by
literally rotating the source code by \(180^\circ\) (more detail on
this below). Despite this restriction on form, we have found Flipper
to be a surprisingly effective language for implementing lossless
compression, and surprisingly enjoyable to use. 

There is a tradition of `reversible' programming languages going back
at least as far as \cite{Lutz86}, and we took particular
inspiration from the pure functional reversible languages rFun
\cite{yokoyama2012,thomsen2015a} and Theseus \cite{james2014}. We believe
that Flipper is the first in this tradition to support dependent
types, and as far as we know the first to be formally verified. We
also think that embedding the reversible language in a well
established host language, with excellent editor integration, makes
writing software in Flipper significantly easier than existing
reversible languages. The power of these features is demonstrated by
our implementation of ANS-based compression in Flipper, which to our
knowledge is the most sophisticated compression method to have been
implemented in a reversible language to date (we plan to publicly release this
soon).

\section{The Flipper language}
The Flipper compiler accepts an
\AgdaField{apply}\AgdaSpace{}%
\AgdaSymbol{:}\AgdaSpace{}%
\AgdaBound{A}\AgdaSpace{}%
\AgdaSymbol{\ensuremath{\rightarrow}}\AgdaSpace{}%
\AgdaBound{B},
expressed as a pattern-matching lambda term with a special
property: rotating the term by \(180^\circ\),
and ``re-righting'' any upside down symbols, results in a valid
term in 
\AgdaBound{B}\AgdaSpace{}%
\AgdaSymbol{\ensuremath{\rightarrow}}\AgdaSpace{}%
\AgdaBound{A}, which we can use as our \AgdaField{unapply}.
\begin{code}[hide]%
\>[0]\AgdaKeyword{open}\AgdaSpace{}%
\AgdaKeyword{import}\AgdaSpace{}%
\AgdaModule{Flipper}\<%
\end{code}

As a first example, consider the `flippable' function which swaps the
elements of a pair:
\begin{code}%
\>[0]\AgdaFunction{pair-swp}\AgdaSpace{}%
\AgdaSymbol{:}\AgdaSpace{}%
\AgdaSymbol{\ensuremath{\forall}}\AgdaSpace{}%
\AgdaSymbol{\{}\AgdaBound{A}\AgdaSpace{}%
\AgdaBound{B}\AgdaSymbol{\}}\AgdaSpace{}%
\AgdaSymbol{\ensuremath{\rightarrow}}\AgdaSpace{}%
\AgdaBound{A}\AgdaSpace{}%
\AgdaOperator{\AgdaFunction{×}}\AgdaSpace{}%
\AgdaBound{B}\AgdaSpace{}%
\AgdaOperator{\AgdaRecord{\ensuremath{\leftrightarrow}}}\AgdaSpace{}%
\AgdaBound{B}\AgdaSpace{}%
\AgdaOperator{\AgdaFunction{×}}\AgdaSpace{}%
\AgdaBound{A}\<%
\\
\>[0]\AgdaFunction{pair-swp}\AgdaSpace{}%
\AgdaSymbol{=}\AgdaSpace{}%
\AgdaFunction{F}\AgdaSpace{}%
\AgdaSymbol{\ensuremath{\lambda}}\AgdaSpace{}%
\AgdaSymbol{\{}\AgdaSpace{}%
\AgdaSymbol{(}\AgdaBound{a}\AgdaSpace{}%
\AgdaOperator{\AgdaInductiveConstructor{,}}\AgdaSpace{}%
\AgdaBound{b}\AgdaSymbol{)}\AgdaSpace{}%
\AgdaSymbol{\ensuremath{\rightarrow}}\AgdaSpace{}%
\AgdaSymbol{(}\AgdaBound{b}\AgdaSpace{}%
\AgdaOperator{\AgdaInductiveConstructor{,}}\AgdaSpace{}%
\AgdaBound{a}\AgdaSymbol{)}\AgdaSpace{}%
\AgdaSymbol{\}}\<%
\end{code}
Here the Flipper compiler, \AgdaFunction{F}, is applied to a lambda
term. Flipping that term gives
\begin{center}
\rotatebox[origin=c]{180}{%
\AgdaSymbol{\ensuremath{\lambda}}\AgdaSpace{}%
\AgdaSymbol{\{}\AgdaSpace{}%
\AgdaSymbol{(}%
\AgdaBound{a}\AgdaSpace{}%
\AgdaInductiveConstructor{,}\AgdaSpace{}%
\AgdaBound{b}%
\AgdaSymbol{)}\AgdaSpace{}%
\AgdaSymbol{\ensuremath{\rightarrow}}\AgdaSpace{}%
\AgdaSymbol{(}%
\AgdaBound{b}\AgdaSpace{}%
\AgdaInductiveConstructor{,}\AgdaSpace{}%
\AgdaBound{a}%
\AgdaSymbol{)}\AgdaSpace{}%
\AgdaSymbol{\}}}.
\end{center}
Rerighting symbols (including the
\AgdaInductiveConstructor{\AgdaUnderscore{},\AgdaUnderscore{}}
constructor/pattern)
and moving the \AgdaSymbol{\ensuremath{\lambda}}
over to the left, we see that in this case
\begin{center}%
\AgdaField{unapply}\AgdaSpace{}%
\AgdaSymbol{\ensuremath{=}}\AgdaSpace{}%
\AgdaSymbol{\ensuremath{\lambda}}\AgdaSpace{}%
\AgdaSymbol{\{}\AgdaSpace{}%
\AgdaSymbol{(}%
\AgdaBound{b}\AgdaSpace{}%
\AgdaInductiveConstructor{,}\AgdaSpace{}%
\AgdaBound{a}%
\AgdaSymbol{)}\AgdaSpace{}%
\AgdaSymbol{\ensuremath{\rightarrow}}\AgdaSpace{}%
\AgdaSymbol{(}%
\AgdaBound{a}\AgdaSpace{}%
\AgdaInductiveConstructor{,}\AgdaSpace{}%
\AgdaBound{b}%
\AgdaSymbol{)}\AgdaSpace{}%
\AgdaSymbol{\}}.%
\end{center}

Note that, in order for the rotated \AgdaField{unapply} to be a valid
Agda function, it is necessary that each bound variable in a flippable
must be used exactly once.  To emphasize the view of \AgdaField{apply}
and \AgdaField{unapply} as different interpretations (or orientations)
of the same program, from here on we will use a more concise,
symmetrical syntax for flippable definitions:
\begin{code}[hide]%
\>[0]\AgdaKeyword{variable}\<%
\\
\>[0][@{}l@{\AgdaIndent{0}}]%
\>[2]\AgdaGeneralizable{A}\AgdaSpace{}%
\AgdaGeneralizable{B}\AgdaSpace{}%
\AgdaGeneralizable{C}\AgdaSpace{}%
\AgdaGeneralizable{X}\AgdaSpace{}%
\AgdaGeneralizable{Z}\AgdaSpace{}%
\AgdaSymbol{:}\AgdaSpace{}%
\AgdaPrimitive{Set}\<%
\end{code}
\begin{code}%
\>[0]\AgdaFunction{pair-swp'}\AgdaSpace{}%
\AgdaSymbol{:}\AgdaSpace{}%
\AgdaGeneralizable{A}\AgdaSpace{}%
\AgdaOperator{\AgdaFunction{×}}\AgdaSpace{}%
\AgdaGeneralizable{B}\AgdaSpace{}%
\AgdaOperator{\AgdaRecord{\ensuremath{\leftrightarrow}}}\AgdaSpace{}%
\AgdaGeneralizable{B}\AgdaSpace{}%
\AgdaOperator{\AgdaFunction{×}}\AgdaSpace{}%
\AgdaGeneralizable{A}\<%
\\
\>[0]\AgdaFunction{pair-swp'}\AgdaSpace{}%
\AgdaSymbol{=}\AgdaSpace{}%
\AgdaFunction{F}\AgdaSpace{}%
\AgdaSymbol{\{}\AgdaSpace{}%
\AgdaSymbol{(}\AgdaBound{a}\AgdaSpace{}%
\AgdaOperator{\AgdaInductiveConstructor{,}}\AgdaSpace{}%
\AgdaBound{b}\AgdaSymbol{)}\AgdaSpace{}%
\AgdaSymbol{\ensuremath{\leftrightarrow}}\AgdaSpace{}%
\AgdaSymbol{(}\AgdaBound{b}\AgdaSpace{}%
\AgdaOperator{\AgdaInductiveConstructor{,}}\AgdaSpace{}%
\AgdaBound{a}\AgdaSymbol{)}\AgdaSpace{}%
\AgdaSymbol{\}}\<%
\end{code}

\begin{figure}[t]
\centering
\setlength{\tabcolsep}{0pt}
\begin{tabular}{lclr}
\(x\)   &       &                                                                      &variables\\
\AgdaInductiveConstructor{\(c\)} &       &                                             &\quad Agda constructors\\
\(T\)   &       &                                                                      &Agda terms\\
\(p\)   &\;::=\;&\(x\) \(\mid\)
         \AgdaSymbol{(}\AgdaInductiveConstructor{\(c\)} \([\) \(p\) \(]\)\AgdaSymbol{)}&patterns\\
\(f\)   &\;::=\;&\AgdaFunction{F} \AgdaSymbol{\{} \(bs\) \AgdaSymbol{\}} \(\mid\) \(T\)&flippables\\
\(bs\)  &\;::=\;&\(b\) \(\mid\) \(b\) \AgdaSymbol{;} \(bs\)                            &branches\\
\(b\)   &\;::=\;&\(p\) \(\AgdaSymbol{\ensuremath\leftrightarrow}\) \(B\)               &\\
\(B\)   &\;::=\;&\(p\) \(\mid\) \(p_1\) \AgdaFunction{\ensuremath\langle} \(f\)
               \AgdaFunction{\ensuremath{\rangle}} \(p_2\)
               \AgdaSymbol{\ensuremath{\leftrightarrow}} \(B\)                         &\\
\end{tabular}
\caption{Grammar of the Flipper language. The notation \([\) \(p\)
\(]\) stands for a list of zero or more patterns. Note that
\emph{infix} Agda constructors, such as
\AgdaInductiveConstructor{\AgdaUnderscore{},\AgdaUnderscore{}},
are also allowed in patterns.}
\end{figure}

New flippables can be built from existing ones: the
syntax
\AgdaBound{a}\AgdaSpace{}%
\AgdaFunction{\ensuremath{\langle}}\AgdaSpace{}%
\AgdaBound{f}\AgdaSpace{}%
\AgdaFunction{\ensuremath\rangle}\AgdaSpace{}%
\AgdaBound{b}\AgdaSpace{}%
\AgdaSymbol{\ensuremath{\leftrightarrow}}\AgdaSpace{}%
\AgdaBound{T}\AgdaSpace{}%
is interpreted in \AgdaField{apply} as ``apply the flippable
\(f\) to the variable \(a\) and bind the result to \(b\) in \(T\)'',
equivalent to the expression
\AgdaKeyword{let}\AgdaSpace{}%
\AgdaBound{b}\AgdaSpace{}%
\AgdaSymbol{=}\AgdaSpace{}%
\AgdaField{apply}\AgdaSpace{}%
\AgdaBound{f}\AgdaSpace{}%
\AgdaBound{a}\AgdaSpace{}%
\AgdaKeyword{in}\AgdaSpace{}%
\AgdaBound{T}. We can, for example, compose two flippables as follows:
\begin{code}%
\>[0]\AgdaOperator{\AgdaFunction{\AgdaUnderscore{}\ensuremath\fatsemi\AgdaUnderscore{}}}\AgdaSpace{}%
\AgdaSymbol{:}\AgdaSpace{}%
\AgdaSymbol{(}\AgdaGeneralizable{A}\AgdaSpace{}%
\AgdaOperator{\AgdaRecord{\ensuremath{\leftrightarrow}}}\AgdaSpace{}%
\AgdaGeneralizable{B}\AgdaSymbol{)}\AgdaSpace{}%
\AgdaSymbol{\ensuremath{\rightarrow}}\AgdaSpace{}%
\AgdaSymbol{(}\AgdaGeneralizable{B}\AgdaSpace{}%
\AgdaOperator{\AgdaRecord{\ensuremath{\leftrightarrow}}}\AgdaSpace{}%
\AgdaGeneralizable{C}\AgdaSymbol{)}\AgdaSpace{}%
\AgdaSymbol{\ensuremath{\rightarrow}}\AgdaSpace{}%
\AgdaGeneralizable{A}\AgdaSpace{}%
\AgdaOperator{\AgdaRecord{\ensuremath{\leftrightarrow}}}\AgdaSpace{}%
\AgdaGeneralizable{C}\<%
\\
\>[0]\AgdaBound{f}\AgdaSpace{}%
\AgdaOperator{\AgdaFunction{\ensuremath\fatsemi}}\AgdaSpace{}%
\AgdaBound{g}\AgdaSpace{}%
\AgdaSymbol{=}\AgdaSpace{}%
\AgdaFunction{F}\AgdaSpace{}%
\>[16]\AgdaSymbol{\{}%
\>[19]\AgdaBound{a}\AgdaSpace{}%
\AgdaSymbol{\ensuremath{\leftrightarrow}}%
\>[25]\AgdaBound{a}%
\>[28]\AgdaOperator{\AgdaFunction{\ensuremath\langle}}\AgdaSpace{}%
\AgdaBound{f}%
\>[34]\AgdaOperator{\AgdaFunction{\ensuremath\rangle}}\AgdaSpace{}%
\AgdaBound{b}%
\>[44]\AgdaSymbol{\ensuremath{\leftrightarrow}}\<%
\\
\>[25]\AgdaBound{b}%
\>[28]\AgdaOperator{\AgdaFunction{\ensuremath\langle}}\AgdaSpace{}%
\AgdaBound{g}%
\>[34]\AgdaOperator{\AgdaFunction{\ensuremath\rangle}}\AgdaSpace{}%
\AgdaBound{c}%
\>[44]\AgdaSymbol{\ensuremath{\leftrightarrow}}\AgdaSpace{}%
\AgdaBound{c}\AgdaSpace{}%
\AgdaSymbol{\}}\<%
\end{code}

To allow for conditional branching, flippable expressions can
contain multiple clauses, corresponding to distinct
input/output patterns. For example, the following flippable swaps the
branches of a sum type:
\begin{code}%
\>[0]\AgdaFunction{sum-swp}\AgdaSpace{}%
\AgdaSymbol{:}\AgdaSpace{}%
\AgdaDatatype{Either}\AgdaSpace{}%
\AgdaGeneralizable{A}\AgdaSpace{}%
\AgdaGeneralizable{B}\AgdaSpace{}%
\AgdaOperator{\AgdaRecord{\ensuremath{\leftrightarrow}}}\AgdaSpace{}%
\AgdaDatatype{Either}\AgdaSpace{}%
\AgdaGeneralizable{B}\AgdaSpace{}%
\AgdaGeneralizable{A}\<%
\\
\>[0]\AgdaFunction{sum-swp}\AgdaSpace{}%
\AgdaSymbol{=}\AgdaSpace{}%
\AgdaFunction{F}\AgdaSpace{}%
\>[15]\AgdaSymbol{\{}%
\>[18]\AgdaSymbol{(}\AgdaInductiveConstructor{left}%
\>[26]\AgdaBound{a}\AgdaSymbol{)}%
\>[30]\AgdaSymbol{\ensuremath{\leftrightarrow}}\AgdaSpace{}%
\AgdaSymbol{(}\AgdaInductiveConstructor{right}%
\>[41]\AgdaBound{a}\AgdaSymbol{)}\<%
\\
\>[15]\AgdaSymbol{;}%
\>[18]\AgdaSymbol{(}\AgdaInductiveConstructor{right}%
\>[26]\AgdaBound{b}\AgdaSymbol{)}%
\>[30]\AgdaSymbol{\ensuremath{\leftrightarrow}}\AgdaSpace{}%
\AgdaSymbol{(}\AgdaInductiveConstructor{left}%
\>[41]\AgdaBound{b}\AgdaSymbol{)}\<%
\\
\>[15]\AgdaSymbol{\}}\<%
\end{code}
In order for a pattern lambda with multiple clauses to be flippable,
the body expressions on the right hand side must partition the output
type: each possible constructor must appear \emph{exactly once}.

Finally, between being bound and being used, variable names are
considered `in scope', and can be freely referred to inside 
\AgdaFunction{\ensuremath{\langle}}\AgdaSpace{}%
\ensuremath\ldots\AgdaSpace{}%
\AgdaFunction{\ensuremath\rangle}\AgdaSpace{}%
(this doesn't count as a `use'), an example is the way that
\AgdaBound{a} is referred to in the flippable
\AgdaFunction{uncurry} combinator:
\begin{code}%
\>[0]\AgdaFunction{uncurry}\AgdaSpace{}%
\AgdaSymbol{:}\AgdaSpace{}%
\AgdaSymbol{(}\AgdaGeneralizable{A}\AgdaSpace{}%
\AgdaSymbol{\ensuremath{\rightarrow}}\AgdaSpace{}%
\AgdaGeneralizable{B}\AgdaSpace{}%
\AgdaOperator{\AgdaRecord{\ensuremath{\leftrightarrow}}}\AgdaSpace{}%
\AgdaGeneralizable{C}\AgdaSymbol{)}\AgdaSpace{}%
\AgdaSymbol{\ensuremath{\rightarrow}}\AgdaSpace{}%
\AgdaGeneralizable{A}\AgdaSpace{}%
\AgdaOperator{\AgdaFunction{×}}\AgdaSpace{}%
\AgdaGeneralizable{B}\AgdaSpace{}%
\AgdaOperator{\AgdaRecord{\ensuremath{\leftrightarrow}}}\AgdaSpace{}%
\AgdaGeneralizable{A}\AgdaSpace{}%
\AgdaOperator{\AgdaFunction{×}}\AgdaSpace{}%
\AgdaGeneralizable{C}\<%
\\
\>[0]\AgdaFunction{uncurry}\AgdaSpace{}%
\AgdaBound{f}\AgdaSpace{}%
\AgdaSymbol{=}\AgdaSpace{}%
\AgdaFunction{F}\AgdaSpace{}%
\AgdaSymbol{\{}\AgdaSpace{}%
\AgdaSymbol{(}\AgdaBound{a}\AgdaSpace{}%
\AgdaOperator{\AgdaInductiveConstructor{,}}\AgdaSpace{}%
\AgdaBound{b}\AgdaSymbol{)}\AgdaSpace{}%
\AgdaSymbol{\ensuremath{\leftrightarrow}}\AgdaSpace{}%
\AgdaBound{b}\AgdaSpace{}%
\AgdaOperator{\AgdaFunction{\ensuremath\langle}}\AgdaSpace{}%
\AgdaBound{f}\AgdaSpace{}%
\AgdaBound{a}\AgdaSpace{}%
\AgdaOperator{\AgdaFunction{\ensuremath\rangle}}\AgdaSpace{}%
\AgdaBound{c}\AgdaSpace{}%
\AgdaSymbol{\ensuremath{\leftrightarrow}}\AgdaSpace{}%
\AgdaSymbol{(}\AgdaBound{a}\AgdaSpace{}%
\AgdaOperator{\AgdaInductiveConstructor{,}}\AgdaSpace{}%
\AgdaBound{c}\AgdaSymbol{)}\AgdaSpace{}%
\AgdaSymbol{\}}\<%
\end{code}

\section{Bits back coding in Flipper}
As a brief example use case, we show how to implement the `bits-back
with ANS' (BB-ANS) method from \cite{townsend2019} in Flipper. This is
not an implementation of ANS itself, but a method for composing
ANS-based codecs in order to encode data using a latent variable
model.
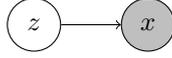
\begin{figure}[h]
\centering
\begin{tikzpicture}
  \tikzstyle{var}=[circle,draw=black,minimum size=7mm]
  \tikzstyle{latent}  =[]
  \tikzstyle{observed}=[fill=gray!50]
  \node at (0,   0) [var,latent]   (z)  {\(z\)};
  \node at (1.5, 0) [var,observed] (x)  {\(x\)};
  \draw[->]  (z) -- (x);
\end{tikzpicture}
\caption{Graphical model with latent variable \(z\) and observed
variable \(x\).}
\end{figure}

We assume a model over data \(x\) where we only have access to the
joint distribution \(P(x, z)\) for some `latent' variable \(z\), and
an approximate posterior \(Q(z\given x)\), but \emph{not} to the
marginal
\begin{equation}
  P(x) = \int_z P(x, z) \;\mathrm{d}z.
\end{equation}

We define a flippable \AgdaFunction{Encoder} type, parameterized by a compressed
message type, \AgdaBound{C}, and the type \AgdaBound{X} of data to be
compressed:
\begin{code}%
\>[0]\AgdaFunction{Encoder}\AgdaSpace{}%
\AgdaSymbol{:}\AgdaSpace{}%
\AgdaPrimitive{Set}\AgdaSpace{}%
\AgdaSymbol{\ensuremath{\rightarrow}}\AgdaSpace{}%
\AgdaPrimitive{Set}\AgdaSpace{}%
\AgdaSymbol{\ensuremath{\rightarrow}}\AgdaSpace{}%
\AgdaPrimitive{Set}\<%
\\
\>[0]\AgdaFunction{Encoder}\AgdaSpace{}%
\AgdaBound{C}\AgdaSpace{}%
\AgdaBound{X}\AgdaSpace{}%
\AgdaSymbol{=}\AgdaSpace{}%
\AgdaBound{C}\AgdaSpace{}%
\AgdaOperator{\AgdaFunction{×}}\AgdaSpace{}%
\AgdaBound{X}\AgdaSpace{}%
\AgdaOperator{\AgdaRecord{\ensuremath{\leftrightarrow}}}\AgdaSpace{}%
\AgdaBound{C}\<%
\end{code}
The \AgdaField{apply} function of an \AgdaFunction{Encoder} accepts
some compressed data in \AgdaBound{C} and a symbol in \AgdaBound{X}
and returns a new element of \AgdaBound{C} from which the original
compressed data and the symbol can both be recovered. An
\AgdaFunction{Encoder}
can be
used to compress a list of elements of \AgdaBound{X}, by (flippably)
folding over the list.

The BB-ANS method can be expressed in Flipper as:
\begin{code}%
\>[0]\AgdaFunction{bb-ans}\AgdaSpace{}%
\AgdaSymbol{:}%
\>[10]\AgdaSymbol{(}\AgdaBound{\ensuremath{P_Z}}%
\>[23]\AgdaSymbol{:}\AgdaSpace{}%
\AgdaFunction{Encoder}\AgdaSpace{}%
\AgdaGeneralizable{C}\AgdaSpace{}%
\AgdaGeneralizable{Z}\AgdaSymbol{)}\<%
\\
\>[10]\AgdaSymbol{(}\AgdaBound{\ensuremath{P_{X\given Z}}}%
\>[23]\AgdaSymbol{:}\AgdaSpace{}%
\AgdaGeneralizable{Z}\AgdaSpace{}%
\AgdaSymbol{\ensuremath{\rightarrow}}\AgdaSpace{}%
\AgdaFunction{Encoder}\AgdaSpace{}%
\AgdaGeneralizable{C}\AgdaSpace{}%
\AgdaGeneralizable{X}\AgdaSymbol{)}\<%
\\
\>[10]\AgdaSymbol{(}\AgdaBound{\ensuremath{Q_{Z\given X}}}%
\>[23]\AgdaSymbol{:}\AgdaSpace{}%
\AgdaGeneralizable{X}\AgdaSpace{}%
\AgdaSymbol{\ensuremath{\rightarrow}}\AgdaSpace{}%
\AgdaFunction{Encoder}\AgdaSpace{}%
\AgdaGeneralizable{C}\AgdaSpace{}%
\AgdaGeneralizable{Z}\AgdaSymbol{)}\<%
\\
\>[10]\AgdaSymbol{\ensuremath{\rightarrow}}\AgdaSpace{}%
\AgdaFunction{Encoder}\AgdaSpace{}%
\AgdaGeneralizable{C}\AgdaSpace{}%
\AgdaGeneralizable{X}\<%
\\
\>[0]\AgdaFunction{bb-ans}\AgdaSpace{}%
\AgdaBound{\ensuremath{P_Z}}\AgdaSpace{}%
\AgdaBound{\ensuremath{P_{X\given Z}}}\AgdaSpace{}%
\AgdaBound{\ensuremath{Q_{Z\given X}}}\AgdaSpace{}%
\AgdaSymbol{=}\AgdaSpace{}%
\AgdaFunction{F}\AgdaSpace{}%
\AgdaSymbol{\{}\AgdaSpace{}%
\AgdaSymbol{(}\AgdaBound{c}\AgdaSpace{}%
\AgdaOperator{\AgdaInductiveConstructor{,}}\AgdaSpace{}%
\AgdaBound{x}\AgdaSymbol{)}\AgdaSpace{}%
\AgdaSymbol{\ensuremath{\leftrightarrow}}%
\>[213I]\AgdaBound{c}%
\>[62]\AgdaOperator{\AgdaFunction{\ensuremath\langle}}\AgdaSpace{}%
\AgdaFunction{flip}\AgdaSpace{}%
\AgdaSymbol{(}\AgdaBound{\ensuremath{Q_{Z\given X}}}\AgdaSpace{}%
\AgdaBound{x}\AgdaSymbol{)}%
\>[85]\AgdaOperator{\AgdaFunction{\ensuremath\rangle}}\AgdaSpace{}%
\AgdaSymbol{(}\AgdaBound{c}\AgdaSpace{}%
\AgdaOperator{\AgdaInductiveConstructor{,}}\AgdaSpace{}%
\AgdaBound{z}\AgdaSymbol{)}%
\>[101]\AgdaSymbol{\ensuremath{\leftrightarrow}}\<%
\\
\>[.][@{}l@{}]\<[213I]%
\>[53]\AgdaSymbol{(}\AgdaBound{c}\AgdaSpace{}%
\AgdaOperator{\AgdaInductiveConstructor{,}}\AgdaSpace{}%
\AgdaBound{x}\AgdaSymbol{)}%
\>[62]\AgdaOperator{\AgdaFunction{\ensuremath\langle}}\AgdaSpace{}%
\AgdaBound{\ensuremath{P_{X\given Z}}}\AgdaSpace{}%
\AgdaBound{z}%
\>[85]\AgdaOperator{\AgdaFunction{\ensuremath\rangle}}\AgdaSpace{}%
\AgdaBound{c}%
\>[101]\AgdaSymbol{\ensuremath{\leftrightarrow}}\<%
\\
\>[53]\AgdaSymbol{(}\AgdaBound{c}\AgdaSpace{}%
\AgdaOperator{\AgdaInductiveConstructor{,}}\AgdaSpace{}%
\AgdaBound{z}\AgdaSymbol{)}%
\>[62]\AgdaOperator{\AgdaFunction{\ensuremath\langle}}\AgdaSpace{}%
\AgdaBound{\ensuremath{P_Z}}%
\>[85]\AgdaOperator{\AgdaFunction{\ensuremath\rangle}}\AgdaSpace{}%
\AgdaBound{c}%
\>[101]\AgdaSymbol{\ensuremath{\leftrightarrow}}\AgdaSpace{}%
\AgdaBound{c}\AgdaSpace{}%
\AgdaSymbol{\}}\<%
\end{code}
This matches the Python implementation presented in \cite[Appendix
C]{townsend2019}, but with only one function required instead of two.
\section{Conclusion}
We have presented the Flipper language, for which we have implemented
a compiler as an Agda macro. Flipper can be used to reduce the amount
of code, and to avoid a class of bugs, when implementing lossless
compression programs. We hope to release Flipper, as well as the full
ANS compression implementation in Flipper, as soon as possible.
\newpage
\section*{Acknowledgements}
Thanks to Wouter Swierstra and Heiko Zimmerman for their advice and
feedback on drafts of this paper.
This publication is part of the project ``neural networks for
efficient storage and communication of information'' (with project
number VI.Veni.212.106) of the research programme
``NWO-Talentprogramma Veni ENW 2021'' which is financed by the Dutch
Research Council (NWO).
\printbibliography
\end{document}